\documentclass[aps,prl,twocolumn,footinbib,bibnotes,preprintnumbers]{revtex4-1} 

\usepackage[utf8]{inputenc}
\usepackage{amsmath,amsfonts,epsfig,color,latexsym}
\usepackage{amssymb}
\usepackage{epsfig,psfrag}
\usepackage{slashed}
\usepackage[normalem]{ulem}

\usepackage{soul,xcolor}
\usepackage{hyperref}
\setstcolor{red}
\usepackage{comment}

\newcommand{\la}{\langle}
\newcommand{\ra}{\rangle}

\usepackage{tikz}

\begin{document}
\newpage
\pagenumbering{arabic}


\title{Coulomb Branch Amplitudes from a Deformed Amplituhedron Geometry}

\author{Nima Arkani-Hamed}
\affiliation{School of Natural Sciences, Institute for Advanced Study,  Princeton, NJ 08540, USA}

\author{Wojciech Flieger}
\affiliation{Max-Planck-Institut f\"{u}r Physik, Werner-Heisenberg-Institut, D-80805 M\"{u}nchen, Germany}

\author{Johannes M. Henn}
\affiliation{Max-Planck-Institut f\"{u}r Physik, Werner-Heisenberg-Institut, D-80805 M\"{u}nchen, Germany}

\author{Anders Schreiber}
\affiliation{Max-Planck-Institut f\"{u}r Physik, Werner-Heisenberg-Institut, D-80805 M\"{u}nchen, Germany}

\author{Jaroslav Trnka}
\affiliation{Center for Quantum Mathematics and Physics (QMAP), Department of Physics, University of California, Davis, CA 95616, USA}
\affiliation{Institute for Particle and Nuclear Physics, Charles University in Prague, Czech Republic}

\preprint{MPP-2023-266}

\begin{abstract}
The Amplituhedron provides, via geometric means, the all-loop integrand of scattering amplitudes in maximally supersymmetric Yang-Mills theory. Unfortunately, dimensional regularization, used conventionally for integration, breaks the beautiful geometric picture. This motivates us to propose a `deformed' Amplituhedron. Focusing on the four-particle amplitude, we introduce two deformation parameters, which can be interpreted as particle masses. We provide evidence that the mass pattern corresponds to a specific choice of vacuum expectation values on the Coulomb branch.
The deformed amplitude is infrared finite, making the answer well-defined in four dimensions. 
Leveraging four-dimensional integration techniques based on differential equations, we compute the amplitude up to two loops. In the limit where the deformation parameters are taken to zero, we recover the known Bern-Dixon-Smirnov amplitude.
In the limit where only one deformation parameter is taken to zero, we find a connection to the angle-dependent cusp anomalous dimension. 
\end{abstract}

\setcounter{tocdepth}{2}
\maketitle
\setcounter{page}{1}

\section{Introduction}

Yang-Mills theory has been the foundation of the Standard Model of Particle Physics for over five decades, describing the fundamental interactions of elementary particles. Despite its importance, the theory remains incompletely understood, and making precision predictions for comparison with experiment is notoriously difficult. 
However, the study of scattering amplitudes has revealed many surprising properties, hidden symmetries and new conceptual approaches that hint at a different formulation of Yang-Mills theory, where Lagrangians and Feynman diagrams do not play a central role. 

In the 't Hooft limit of a large number of colors, only planar Feynman diagrams contribute. 
The latter may be seen as approximating a two-dimensional surface, hinting at a geometric formulation of the theory.  The maximally supersymmetric Yang-Mills theory (sYM) is an ideal testing ground for such novel ideas. In the AdS/CFT correspondence, this picture is made more concrete: the scattering amplitudes at strong coupling are described by the volume of certain minimal surfaces \cite{Alday:2007hr}.

How does such a geometric picture emerge from perturbation theory? 
The Amplituhedron \cite{Arkani-Hamed:2013jha,Arkani-Hamed:2017vfh}  provides a novel geometric framework for defining the all-loop integrand of the planar sYM amplitudes. The latter is obtained as the canonical differential form 
on a certain geometric space, defined by a set of inequalities that involve the particle kinematics.
Important work by mathematicians and physicists has been devoted to understanding, in detail, the geometry of the space \cite{Karp:2016uax,Karp:2017ouj,Lukowski:2020dpn,Parisi:2021oql,Even-Zohar:2021sec,Blot:2022geq},
\cite{Arkani-Hamed:2013kca,Franco:2014csa,Arkani-Hamed:2014dca,Arkani-Hamed:2018rsk,Herrmann:2020qlt,Dian:2022tpf,Herrmann:2022nkh,Arkani-Hamed:2021iya,Damgaard:2019ztj,Ferro:2022abq,He:2023rou}. 

We expect that this improved understanding of the geometric origin of the integrand will be important for determining the final answer after integrating over Minkowski space. 
Unfortunately, the fact that massless particles have infrared divergences requires the use of a regulator. The standard method, dimensional regularization, breaks the beautiful geometric picture. This motivates us to propose a `deformed' Amplituhedron. Focusing on the four-particle amplitude, in this paper we introduce certain shifts in the kinematics. The latter are somewhat reminiscent of, but different from, BCFW shifts \cite{Britto:2005fq}. As a result, we obtain a deformed amplitude that depends on two deformation parameters and that is well-defined in four dimensions. In a nutshell, the input is the four-particle scattering kinematics; after working out the form and performing the integrations, the output is a finite amplitude, at a given loop order. 
Our formula exactly matches certain  Coulomb branch amplitudes.


\section{Four-point Deformed Amplituhedron}
\label{sec:deformedgeometry}

\begin{figure}[t]
    \centering
    \includegraphics[scale=0.4]{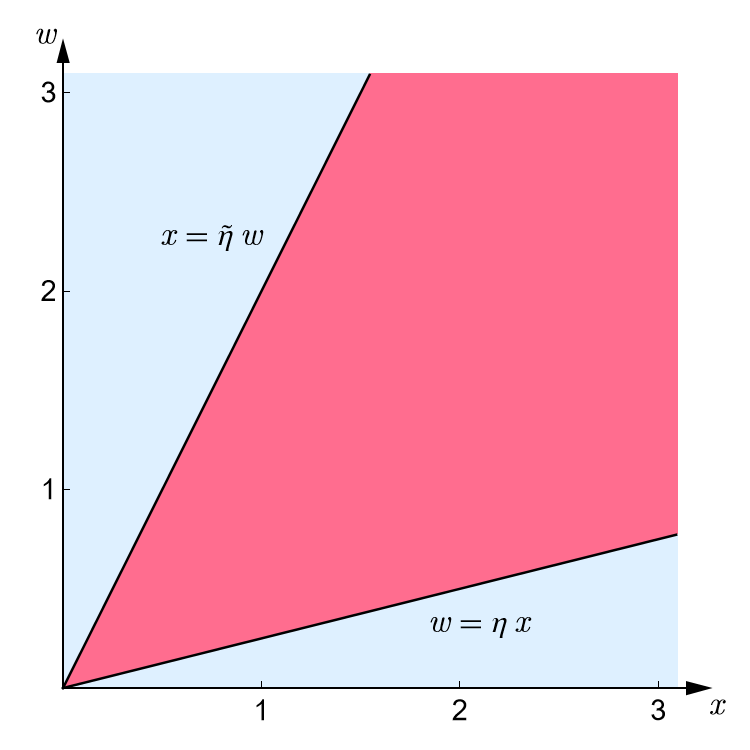}
\caption{
\label{fig:deformationetaetabar}
Illustration of the undeformed Amplituhedron geometry (first quadrant, in light blue), for $\eta= \tilde{\eta}=0$, and of the
deformed geometry (wedge, in red) 
for $\eta=\frac{1}{4}$, $\tilde{\eta}=\frac{1}{2}$. }
\end{figure}

The definition of the four-point Amplituhedron geometry consists of two components: four external momentum twistors $Z_1$, $Z_2$, $Z_3$, $Z_4$ and $L$ lines $\{(AB)_i\}$. The former satisfy the positivity condition $\la 1234\ra \equiv \det(Z_1 Z_2 Z_3 Z_4) {>} 0$, and can be considered as fixed points in the momentum twistor space. The Amplituhedron is then a configuration of lines $\{(AB)_i\}$ satisfying the following inequalities: 
\begin{align}\label{pos1}
\begin{split}
& \la (AB)_i 12\ra{>}0,\, \la (AB)_i23\ra{>}0,\,
 \la (AB)_i34\ra{>}0,\,  \\ &  \la (AB)_i14\ra{>}0 \,, \la (AB)_i13\ra{<}0,\, \la (AB)_i24\ra{<}0 \,,
\end{split}
\end{align}
with the last two inequalities following from the
sign flip condition \cite{Arkani-Hamed:2017vfh}.
%
%
Additional positivity conditions are imposed between each pair of lines $(AB)_i$ and $(AB)_j$,
\begin{equation}
    \la (AB)_i(AB)_j\ra > 0 \,.\label{pos3}
\end{equation}
The external momentum twistors 
form four lines $X_1\equiv Z_1Z_2$, $X_2\equiv Z_2Z_3$, $X_3\equiv Z_3Z_4$, $X_4 \equiv Z_1Z_4$, 
which correspond to massless propagators, by virtue of $\la X_iX_i\ra =0$. 
Adjacent pairs of lines intersect,
\begin{equation}
\la X_1X_2\ra = \la X_2X_3\ra = \la X_3X_4\ra = \la X_4X_1\ra = 0 \,, \label{adjacent}
\end{equation}
To obtain the deformed Amplituhedron, one shifts the external kinematics $X_i \rightarrow \widehat{X}_i$. This is done while preserving (\ref{adjacent}) but allowing $\la \widehat{X}_i\widehat{X}_i\ra \neq 0$. 
We define
\begin{align}
   & \widehat{X}_1 = X_1 - \rho X_3, \quad \widehat{X}_3 = X_3 - \tilde{\rho} X_1\,, \\
   & \widehat{X}_2 = X_2 - \eta X_4, \quad \widehat{X}_4 = X_4 - \tilde{\eta} X_2\,.
\end{align}
We now replace ${X}_i \to \widehat{X}_i$ in the first four inequalities of  (\ref{pos1}). Other inequalities remain unchanged.
Using the following parametrization of the $(AB)_i$ lines,
\begin{equation}
    Z_{A_i} = Z_1 + x_i Z_2 - y_i Z_4,\quad Z_{B_i} = Z_3 + z_i Z_2 + w_i Z_4 \,,
\end{equation}
the original conditions (\ref{pos1}), imply $x_i,y_i,z_i,w_i>0$, while instead in the shifted case we have (cf. Fig.~\ref{fig:deformationetaetabar}),
\begin{align}\label{ineq1}
  &  y_i > \rho z_i > 0, \quad\,\, z_i > \tilde{\rho} y_i > 0 \, , \\
  & \hspace{-0.06cm}  w_i > \eta x_i > 0 , \quad x_i > \tilde{\eta} w_i > 0 \, ,  \label{ineq2}
\end{align}
where we further impose conditions on shift parameters $\rho,\tilde{\rho},\eta,\tilde{\eta}>0$,
such that $\rho\tilde{\rho}<1$ and $\eta\tilde{\eta}<1$.

The boundary structure of the deformed Amplituhedron happens to be much simpler than of the original space. Thanks to the deformation parameters, all collinear configurations are removed as boundaries. As a consequence, the number of boundaries of the highest codimension (leading singularities) is only two rather than six: only the configurations with $(AB)_i =13$ and $(AB)_i =24$ remain, while all other soft-collinear configurations $(AB)_i = k\,k{+}1$ are not present. At one-loop, the undeformed Amplituhedron is a positive Grassmannian $G_+(2,4)$ with a number of boundaries of various co-dimensions, $(4,10,12,6)$, and Euler characteristic $\chi = +1$. For example, the codimension-two boundaries correspond to four double cuts $\la AB12\ra{=}\la AB23\ra{=}0$ (and cyclic), each having two solutions -- either $AB$ passing $Z_2$ or being in the plane $(Z_1Z_2Z_3)$, and two double cuts $\la AB12\ra{=}\la AB34\ra{=}0$ and $\la AB23\ra{=}\la AB14\ra{=}0$, totaling 10 codimension-two boundaries. In the deformed case any double cut, $\la AB\hat{X}_i\ra = \la AB\hat{X}_j\ra = 0$, has only one solution, hence we have only six codimension-two boundaries. The number of boundaries of various co-dimensions is now $(4,6,4,2)$, which demonstrates the simplicity of the geometry. The Euler characteristic is again $\chi = +1$. 

\section{Two-loop integrand}

At two-loops, the geometric space is more involved and the direct triangulation, though possible, is more difficult. Therefore here we determine the form by making the following ansatz, 
\begin{equation}
  \Omega^{(2)} =\frac{d\mu_{AB}d\mu_{CD}\cdot {\cal N}(AB,CD,Z_i,\rho,\tilde{\rho},\eta,\tilde{\eta})}{ \bigg\{\hspace{-0.1cm}\begin{array}{c}
    \la AB\hat{X}_1\ra\la AB\hat{X}_2\ra\la AB\hat{X}_3\ra\la AB\hat{X}_4\ra\la ABCD\ra \\
\la CD\hat{X}_1\ra\la CD\hat{X}_2\ra\la CD\hat{X}_3\ra\la CD\hat{X}_4\ra \end{array}\hspace{-0.1cm}\bigg\}} \, , 
\end{equation}
which is based on the known denominator structure and impose consistency conditions on the numerator. One of these conditions is the absence of ``non-planar" cuts, which violate the $\la ABCD\ra>0$ condition. For example, we can cut following propagators,
\begin{equation}
\la AB\hat{X}_1\ra = \la AB\hat{X}_2\ra = \la CD\hat{X}_1\ra = \la CD\hat{X}_2\ra = 0 \, , 
\end{equation}
which fixes $y_1=\rho z_1$, $y_2=\rho z_2$, $w_1 = \eta x_1$, $w_2=\eta x_2$. Then evaluating $\la ABCD\ra$ in this parametrization gives
\begin{equation}
\la ABCD\ra = -\rho (z_1-z_2)^2 - \eta (x_1-x_2)^2 \, . 
\end{equation}
This expression is manifestly negative, which is in tension with the $\la ABCD\ra>0$ condition. Hence this cut is unphysical and the numerator must vanish when evaluated on it. Imposing all such conditions the ansatz reduces to the sum of deformed double-box integrals (in  analogy with the non-deformed case),
\begin{align}
\Omega^{(a)}_{\rm db} &= - \frac{d\mu_{AB}d\mu_{CD}\cdot  \la \hat{X}_1 \hat{X}_3 \ra^2  \la \hat{X}_2 \hat{X}_4 \ra }{ \bigg\{ \hspace{-0.1cm} \begin{array}{c}\la AB\hat{X}_1\ra\la AB\hat{X}_2\ra\la AB\hat{X}_3\ra\la ABCD\ra\\ \la CD\hat{X}_3\ra\la CD\hat{X}_4\ra\la CD\hat{X}_1\ra\end{array} \hspace{-0.1cm} \bigg\}},\\
\Omega^{(b)}_{\rm db} &= - \frac{d\mu_{AB}d\mu_{CD}\cdot  \la \hat{X}_1 \hat{X}_3 \ra  \la \hat{X}_2 \hat{X}_4 \ra^2}{\bigg\{
\hspace{-0.1cm} \begin{array}{c}\la AB\hat{X}_2\ra\la AB\hat{X}_3\ra\la AB\hat{X}_4\ra\la ABCD\ra\\ \la CD\hat{X}_4\ra\la CD\hat{X}_1\ra\la CD\hat{X}_2\ra\end{array} \hspace{-0.1cm} \bigg\}}.
\end{align}
and $\Omega^{(2)}$ is a symmetrized sum in $AB\leftrightarrow CD$,
\begin{equation}
    \Omega^{(2)} = n^{(2a)}\Omega^{(a)}_{\rm db} + n^{(2b)}\Omega^{(b)}_{\rm db} + (AB\leftrightarrow CD)
\end{equation}
Unlike in the massless case there is no cancellation of spurious cuts between $\Omega^{(a)}_{\rm db}$ and $\Omega^{(b)}_{\rm db}$, and their coefficients $n^{(2a)}$ and $n^{(2b)}$ are unrelated. 

%
%

\section{Amplitudes from Deformed Amplituhedron}
\label{sec:deformedgeometry2}

The Amplituhedron defines the planar $L$-loop integrand $\Omega^{(L)}$ at all orders in perturbation theory. More concretely, working with the amplitude, normalized by its tree-level contribution, we have
\begin{align}
M = 1 + g^2 M^{(1)} + g^4 M^{(2)} + {\cal{O}}(g^6) \,,
\end{align}
with $g^2 = g_{\rm YM}^2 N_c /(16 \pi^2)$, and 
\begin{align}\label{eqdefMfromomega}
    M^{(L)} = \int \Omega^{(L)}\,.
\end{align}
The $L$-loop integrand $\Omega^{(L)}$ is the form with logarithmic singularities on the boundaries of the deformed Amplituhedron space. 

The one-loop form associated to the deformed Amplituhedron is found to be 
\begin{equation}
    \Omega^{(1)} = - n^{(1)} \frac{d\mu_{AB} 
         \la \hat{X}_1 \hat{X}_3 \ra   \la \hat{X}_2 \hat{X}_4 \ra 
    }
    {\la AB\hat{X}_1\ra\la AB\hat{X}_2\ra\la AB\hat{X}_3\ra\la AB\hat{X}_4\ra} \,, \label{oneloopintegrand}
\end{equation}
where $d\mu_{AB} = \la ABd^2A\ra\la ABd^2B\ra$ is measure 
in  momentum twistor space, normalized such that \cite{Arkani-Hamed:2010pyv}
\begin{align}\label{defmeasure}
\int \frac{d\mu_{AB}}{\langle AB Q \rangle^4}  =\frac{1}{\Gamma(4)} \frac{1}{   \left( \frac{1}{2} \langle Q Q \rangle \right) ^2} \,.
\end{align}
The normalization $n^{(1)}$ is not determined by the Amplituhedron, and it is related to the value of the leading singularity -- in the undeformed case the latter is equal to 1. Note that in general, $n^{(1)}$ is a function of the deformation parameters $\rho,\tilde{\rho},\eta,\tilde{\eta}$.

Thanks to 
dual conformal symmetry, the
integrated amplitude $M$ depends on two invariants only, which are
%

\begin{align}\label{crossratiosintermsofparameters}
    u = \frac{\la \hat{X}_1 \hat{X}_3 \ra^2}{\hat{X}_1^2 \hat{X}_3^2} 
    \,,\qquad
      v = \frac{\la \hat{X}_2  \hat{X}_4 \ra^2}{ \hat{X}_2^2 \hat{X}_4^2} 
       \,,
\end{align}
which in terms of deformation parameters are given by
$( 1+ \rho \tilde{\rho})^2/(4 \rho \tilde{\rho})$ and $( 1+ \eta \tilde{\eta})^2/(4 \eta \tilde{\eta})$, respectively.

\section{Comparison to Coulomb-branch amplitudes}
\label{sec:comparisoncoulomb}

%
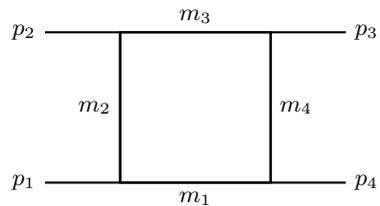
\begin{figure}[t]
\begin{tikzpicture}
    \draw[thick] (1,0) node[left] {$p_1$} -- (2,0);
    \draw[thick] (1,2) node[left] {$p_2$} -- (2,2);
    \draw[thick] (4,0) -- (5,0) node[right] {$p_4$};
    \draw[thick] (4,2) -- (5,2) node[right] {$p_3$};
    %
    \draw[thick] (2,0) -- (2,2) node[midway,left] {$m_2$};
    \draw[thick] (2,0) -- (4,0) node[midway,below] {$m_1$};
    \draw[thick] (2,2) -- (4,2) node[midway,above] {$m_3$};
    \draw[thick] (4,0) -- (4,2) node[midway,right] {$m_4$};
    %
    \draw[thick] (2,0) -- (4,0) -- (4,2) -- (2,2) -- cycle;
   %
\end{tikzpicture}
\caption{
\label{fig:momentumkinematics}
One-loop box integral in the deformed Amplituhedron kinematics.
The on-shell conditions read $p_i^2 = m_i^2 + m_{i+1}^2$, with $m_{5} \equiv m_1$.
}
\end{figure}
Interestingly, our deformation leads to massive particles, without the need of introducing an infinity twistor. 
Translating the kinematics of the deformed Amplituhedron from momentum twistor space to ordinary momentum space, we find
that the amplitude corresponds to scattering of particles, as shown in Figure \ref{fig:momentumkinematics}.
The masses are simple functions of the deformation parameters, and vanish when the latter are sent to zero.
%
%
The dual conformal invariants are related to the kinematics as follows,
\begin{align}
u = \frac{(-s+ m_1^2+m_3^2)^2}{4 m_1^2 m_3^2}\,,\quad
v = \frac{(-t+ m_2^2+m_4^2)^2}{4 m_2^2 m_4^2}\,,
\end{align}
where $s=(p_1+p_2)^2$, $t=(p_2+p_3)^2$.
Thanks to the masses, the amplitude is infrared finite. This finiteness property is built into our geometric setup: it can be seen as a consequence of the absence of collinear and soft-collinear regions in the boundary structure of the deformed Amplituhedron. 

Our setup is reminiscent of the AdS-inspired Coulomb branch amplitudes \cite{Alday:2009zm,Correa:2012nk}, where one considers a Higgs mechanism $U(4+N) \to U(4) \times U(N)$. 
This is done by choosing vacuum expectation values in the $U(4)$ part as follows, 
\begin{align}\label{newvevs}
\langle{\Phi}\rangle = \frac{1}{g} {\rm diag}(m_1 \vec{n}_1, m_2 \vec{n}_2, m_3 \vec{n}_3,m_4 \vec{n}_4)\,,
\end{align}
where $\vec{n}_i$ are $SO(6)$ vectors. 
In \cite{Alday:2009zm}, all $\vec{n}_i$ were chosen to be equal, which lead to a certain mass configuration. Instead, we find that we can match the mass configuration above by choosing $\vec{n}_1 = -\vec{n}_3$, $\vec{n}_2 = -\vec{n}_4$, with $\vec{n}_1\cdot \vec{n}_2 = 0$.\footnote{We note that it is also possible to switch on further angles, such that $\vec{n}_1 \cdot \vec{n}_3 = - \cos \theta_{13}$ and $\vec{n}_2 \cdot \vec{n}_4 = - \cos \theta_{24}$.}
We conjecture that the deformed Amplituhedron describes these Coulomb branch amplitudes.
If this conjecture is true, then we can use unitarity of the Coulomb branch amplitudes to determine the relative coefficients of the positive geometries.
Indeed, performing a one-loop computation as in Appendix B of  \cite{Alday:2009zm}, we find that $n^{(1)}=1$, and likewise one may fix $n^{(2a)} = n^{(2b)} =1$ from generalized unitarity.

\section{Integrated 
one- and two-loop 
amplitudes}
\label{sec:twoloop}

For carrying out the integration, we find it convenient to parametrize the kinematics as follows,
\begin{align}\label{defxy}
\rho = \tilde{\rho} = -x \,,\quad \eta= \tilde{\eta} = -y \,.
\end{align}
with $0<x<1, 0<y<1$.
This corresponds to a subset of the Euclidean kinematic region, where the answer is real-valued \footnote{Note that although the cross-ratios are seemingly independent e.g. under $x \leftrightarrow -x$, this transformation requires analytic continuation. See e.g. Duplancic, Nizic, Eur.Phys.J.C 24 (2002) for a related discussion.}.
Note that this is different from the Amplituhedron region.
The cross-ratios of eq. (\ref{crossratiosintermsofparameters}) then take the form
\begin{align}\label{cross-ratios}
    u =  \frac{1}{4} \left( x+\frac{1}{x} \right)^2
    \,,\qquad
      v =  \frac{1}{4} \left( y+\frac{1}{y} \right)^2
       \,.
\end{align}

Starting from eqs. (\ref{eqdefMfromomega}) and 
(\ref{oneloopintegrand}), one may obtain a parameter integral representation for $M^{(1)}$ 
by introducing Feynman parameters, and then using the basic identity (\ref{defmeasure}),
 with the result
\begin{align}
M^{(1)}(x,y) &=  
- {
\frac{(1-x) (1-y)} {(1+x) (1+y)}} \int_{0}^{\infty}\frac{d^{4} \alpha}{GL(1)} \times \\
&\hspace{-2cm} 
\frac{(1-x^{2})\, (1-y^{2})}{[(\alpha_{1}x+\alpha_{3})(\alpha_{1}+\alpha_{3}x)+(\alpha_{2}y+\alpha_{4})(\alpha_{2}+\alpha_{4}y)]^{2}}  \,.\nonumber       
\label{feyn_para} 
\end{align}
Carrying out the integration yields a surprisingly simple answer,
\begin{align}
M^{(1)}(x,y) = -2  {
\frac{(1-x) (1-y)} {(1+x) (1+y)}}  \log (x) \log (y) \,.
\end{align}
At two loops, 
we have 
\begin{align}
M^{(2)} (x,y) =   I_{\rm db}(x,y) + I_{\rm db}(y,x)  \,,
\end{align}
where
\begin{align}
I_{\rm db}(x,y)  = \int \Omega^{(a)}_{\rm db}\,,\qquad 
I_{\rm db}(y,x)  = \int \Omega^{(b)}_{\rm db}
\,.
\end{align}
Leveraging four-dimensional differential equation techniques \cite{Caron-Huot:2014lda} we find that, remarkably, the answer for the double box can be written in just a few lines,
\begin{align}
I_{\rm db}(x,y) =&
 {
 \frac{(1-x)^2 (1-y)} {(1+x)^2 (1+y)}} 
\Big[   J_{3}(x^2) \log (y^2) \notag \\
 & -Q(y^2) + \frac{1}{2} Q\left(\frac{y^2}{x^2}\right) + \frac{1}{2} Q(x^2 y^2) \Big] 
\,,
\end{align}
with 
\begin{align}
   \hspace{-0.2em} Q(z) &= 
   3 \text{Li}_4 (z) -3 \log(z) \text{Li}_3 (z) + \frac{3}{2} \log^2 (z) \text{Li}_2 (z) \nonumber \\
   &+ \frac{1}{2} \log^3 (z) \log(1-z) 
    + \frac{3 \pi^4}{10} +\frac{\pi^2}{4} \log^2 \left( z \right) \nonumber \\ & +\frac{3}{16} \log^4 \left( z \right) + \log^2 \left( z \right) {\rm Li}_{2}(1-z) + 4 \pi^2 {\rm Li}_{2} \left(-\sqrt{z} \right) \nonumber \\ 
    &- \log \left( z \right)  {\rm Li}_{3}\left( 1- \frac{1}{z} \right)  - \log \left(z\right) {\rm Li}_{3}\left( 1- {z} \right) \, ,   
\end{align}
and
\begin{align}
\begin{split}
J_{3}(z) &= \frac{1}{4} \log^3 \left(z \right) + \log \left( z \right)  {\rm Li}_{2}\left( 1- {z} \right)   \\
&-2  {\rm Li}_{3}\left( 1- {z} \right) -2  {\rm Li}_{3}\left( 1- \frac{1}{z} \right) \,.
\end{split}
\end{align}
$Q$ and $J_3$ are real-valued for positive arguments.
As in the famous result for the six-gluon two-loop remainder function in sYM \cite{Goncharov:2010jf}, only classical polylogarithms are
needed. 
In contrast, the two-loop four-particle amplitude in the (slightly different) Coulomb-branch setup of \cite{Alday:2009zm} is more complicated \cite{Caron-Huot:2014lda} .

We also note that the amplitude up to two loops has the following ten-letter symbol alphabet,
\begin{align}\label{eq:alphabet}
\begin{split}
{\cal A} = & \{ x,1 \pm x, y,1 \pm y, x\pm y, 1 \pm x y   \}\,.
\end{split}
\end{align}
It is interesting to study the behavior of $M$ near singular kinematic configurations, i.e. where one or more terms in eq. (\ref{eq:alphabet}) vanish. 
We now present exact formulas for the leading behavior in two such limits.

\section{Exact results in kinematic limits}
{\it High-energy limit.} This corresponds to $s,t \to \infty$, or equivalently, $x,y \to 0$, keeping $x/y = t/s$ fixed.
This limit where the deformation parameters are taken to zero allows us to connect to 
the massless, infrared-divergent amplitude. 
We conjecture that the following exact formula holds in the high-energy limit, 
\begin{align}\label{Mconjecturehighenergy}
\begin{split}
 \log M \stackrel{x,y \to 0}{=} & -\frac{1}{2} \Gamma_{\rm cusp}(g)  \log x \log y \\
 &\hspace{-0cm} + \Gamma_{\rm collinear}(g) \left( \log x + \log y \right) +  {C}(g) \,,
\end{split}
\end{align}
which is analogous to formulas in dimensional regularization \cite{Bern:2005iz,Drummond:2007au} and on the Coulomb branch \cite{Alday:2009zm}.
Here $\Gamma_{\rm cusp} = 4 g^2 - 8 \zeta_2 g^4 + {\cal O}(g^6)$ is the light-like cusp anomalous dimension, which is famously known from integrability \cite{Beisert:2006ez}.
Comparing to our perturbative results, we find that (\ref{Mconjecturehighenergy}) holds with
the collinear anomalous dimension $\Gamma_{\rm collinear}(g)=-4 \zeta_3 g^4+ {\cal O}(g^6)$ and $ {C}(g)=-3/10 \pi^4 g^4+ {\cal O}(g^6)$. The latter two are both regularization-scheme dependent \cite{Henn:2011by}.

{\it Regge limit $t \to \infty$.} 
This corresponds to $y\to0$, keeping $x$ fixed.
Thanks to dual conformal symmetry, this may be equivalently viewed as the small mass $m_{1} \to 0$ limit. 
Therefore we expect this limit to be governed by eikonal physics  \cite{Henn:2010bk}, and in particular by the anomalous dimension of a cusped Wilson loop, which to two loops is given by \cite{Drukker:2011za}
\begin{align}
\Gamma_{\rm cusp}(\phi,\theta; g) =&  g^2  \xi (- 2 \log x)  + \\
& \hspace{-2cm} +  g^4 \Bigg\{ \xi \frac{4}{3} \log x  \left( \pi^2 + \log^2 x  \right)  
+ \xi^2 \Bigg[ 4 \text{Li}_3\left(x^2\right)   \nonumber \\
&  \hspace{-1.8cm} -4 \text{Li}_2\left(x^2\right) \log (x)-\frac{4}{3} \log ^3(x)-\frac{2}{3} \pi ^2
   \log (x)-4 \zeta_3 \Bigg]  \Bigg\} \,. \nonumber
\end{align} 
Here the Euclidean cusp angle $\phi$ is related to $x$ via $x=e^{i\phi}$, and the second parameter $\theta$ parametrizes the Wilson loop's coupling to scalars,
via the combination
\begin{align}
\xi = \frac{\cos \theta - \cos \phi}{i \sin \phi } 
= \frac{1+x^2-2 x \cos \theta }{1-x^2}\,.
\end{align}
The choice of SO(6) vectors we made below eq. (\ref{newvevs}) suggests that here $\theta=0$, which implies $\xi = (1-x)/(1+x)$.
Indeed, we find that the leading terms in the Regge limit are given by
\begin{align}\label{eq:limitMRegge}
  M(x=e^{i \phi},y)  \stackrel{y \to 0}{=} & r(\phi, 0 ; g) \, y^{\Gamma_{{\rm cusp}}(\phi; 0; g)}+ {\cal O}({y^0})  \,,
\end{align}
where $r(\phi, 0 ; g)$ being the finite part in the Regge limit.
It is interesting to note that by making a more general choice of $SO(6)$ vectors in eq. (\ref{newvevs}), such that $\vec{n}_1 \cdot \vec{n}_3 = - \cos \theta$, we can match the full $\xi$ dependence.

\section{Summary and Outlook}
\label{sec:outlook}

We generalized the four-particle Amplituhedron geometry of planar sYM such that the amplitude $M(x,y)$ is infrared finite and depends on two dual conformal parameters $x,y$.  
The finiteness is due to massive propagators.
The mass configuration is different from the Coulomb branch setup of \cite{Alday:2009zm}, but we found that a slightly different pattern of vacuum expectation values matches our geometric setup. It remains to be proven that the deformed Amplituhedron not only matches the kinematics, but gives exactly those Coulomb branch amplitudes.

In the original Amplituhedron, fixing the integrand involved the assumption of normalizing the canonical forms to have unit (constant) leading singularities on certain boundaries. Instead, the Coulomb branch amplitudes in general have kinematic-dependent leading singularities. This means that a generalization is needed of how we think about canonical forms. In this Letter, we fixed those normalizations by comparing against generalized cuts of Coulomb branch amplitudes. One may ask -- are there other ways of fixing the answer that do not refer to field theory at all?

Given the two-loop integrand obtained in this way, we computed the analytic result for $M(x,y)$ up to two loops, and conjectured exact formulas both in the high-energy as well as the Regge limit.
Given the relatively simple symbol alphabet that we uncovered, it may be possible to bootstrap higher-loop results, given sufficient physical input from limiting configurations and analytic properties (for a review see \cite{Caron-Huot:2020bkp}).

We expect that the new setup will lead to substantial progress in making the connection between geometry and the integrated functions, benefiting from recent work in mathematics \cite{Even-Zohar:2021sec}. This could lead to a geometry-based differential equations method, with broader applications. 

Our novel starting point also gives completely new ways of studying physical limits, 
and to potentially obtain exact results. Thanks to the simpler geometric configurations in the limit, obtaining the canonical form and integrating it should be easier. 
One particularly interesting example is the $\phi \to 0$ limit of the angle-dependent cusp anomalous dimension, whose first term $\Gamma_{\rm cusp}(\phi,\theta=0;g) = - \phi^2 B(g) + {\cal O}(\phi^4)$ is the exactly-known Bremsstrahlung function \cite{Correa:2012at}.

\section*{Acknowledgments}

We thank Simon Caron-Huot, Diego Correa and Martín Lagares for  discussions. This research received funding from the European Research Council (ERC) under the European Union's Horizon 2020 research and innovation programme (grant agreement No 725110), {\it Novel structures in scattering amplitudes},  GA\v{C}R 21-26574S, DOE grants No. SC0009999 and No. SC0009988, and the funds of the University of California.
 
 \bibliographystyle{apsrev4-1} 
\bibliography{amplituhedron.bib}

\end{document}